\documentclass[aps,prd,twocolumn,showpacs,groupedaddress,floatfix]{revtex4}

\usepackage{graphicx}

\begin{document}



\title{Can the universe afford inflation?}


\author{Andreas Albrecht}
\author{ Lorenzo Sorbo }
\affiliation{Department of Physics, UC Davis,
One Shields Avenue, Davis CA, USA 95616}


\date{\today}

\begin{abstract}
Cosmic inflation is envisioned as the ``most
likely'' start for the observed universe. To give substance to this
claim, a framework is needed in
which inflation can compete with other scenarios and the relative likelihood
of all scenarios can be quantified.  The most concrete scheme to date
for performing such a comparison shows inflation to be strongly
disfavored. We analyze the source of this failure for inflation and
present an alternative calculation, based on more traditional
semiclassical methods, that results in inflation being exponentially
favored. We argue that reconciling the two contrasting approaches
presents interesting fundamental challenges, and is likely to have a
major impact on ideas about the early universe.

\end{abstract}


\maketitle


\section{Introduction}

Over the last twenty years cosmic inflation theory \cite{infl} 
has survived
extensive theoretical and observational scrutiny and has come to be
seen as the leading theory of the origin of the universe (see for
example \cite{Peiris:2003ff}).  There are
still a number of fundamental open questions for cosmic
inflation. Some of these questions are sufficiently significant that
their 
resolution could severely undermine cosmic inflation
as a theory of cosmic origins.

One of these open questions is the topic of this paper: How inflation
itself got started. The very first papers on inflation treated
inflation as a small modification to the big bang, a particular phase
in the evolution of a Friedmann Robertson Walker (FRW) universe that
started (as usual) with the initial FRW singularity.  But very soon
\cite{lindechaotic,Vilenkin:de,Hartle:1983ai,Farhi:1986ty,Linde:1991sk} another
view developed   
that cosmic inflation should be regarded as a mechanism
which can create the standard big bang (SBB) 
cosmology\footnote{ 
Here SBB cosmology refers to cosmology that gives
the standard FRW  picture of the observed universe back to some early
time, at which point it could match on to reheating at the end of
inflation or some other path out of a larger ``meta-universe''. }
out of a 
fluctuation originating in some ``meta universe''.  By
``meta-universe'' we refer to whatever theory one has to describe (and
attach probabilities to) the range of fluctuations which might
possibly create a big bang universe (and there are a variety of
proposals for this). In this newer picture, the
pre-inflation cosmological evolution is given by the random
fluctuations in the meta-universe, some of which gives rise to
inflation.  

One of the main attractions of inflation has been
that it offers an account of the origin of the universe that seems ``more
natural'' or ``more likely'' than the standard big bang taken on its
own.  This perception is typically based on rather vague but
intuitively reasonable arguments about the attractor nature of
inflationary dynamics and about fine tuning of initial conditions.
The only real proposals to treat this aspect of inflation in a more
rigorous way are ones that place inflation in direct competition with
other mechanisms for creating the big bang cosmology in which we
live.  If one can actually assign relative
probabilities to the observed big bang universe fluctuating out of the
meta-universe through different ``channels'', some including and
others not including an inflationary phase, one can then
quantify the degree to which inflation really is more likely to
describe the  history of the region of the universe we observe. This
approach has been emphasized recently in 
\cite{Garriga:1997ef,Dyson:2002pf,Hawking:2002af,Turok:2000bt,Turok:2002yq,Coule:2002zb,Albrecht:2002uz,Hawking:2003bf}.         

In \cite{Dyson:2002pf} Dyson, Kleban, and Susskind (DKS) provide what
is probably 
the most concrete calculation of this sort to date. Their scheme is 
particularly attractive because it defines the meta-universe as an
equilibrium state, and uses statistical mechanics to evaluate
different probabilities of fluctuations out of equilibrium. Thus
dynamics, rather than any ad hoc assumptions about ``state of the
universe'' determine the properties of the
meta-universe \footnote{Banks et al raise the interesting question of
  whether such a system is even well defined quantum mechanically
  given that no quantum measurement can be stable over times longer
  than the recurrence time \cite{Banks:2002wr}. Here we take the view
  that quantum measurement is an emergent process which need not be
  eternally stable for a fundamental quantum formulation of a system
  to be valid.}.  As
discussed in \cite{Albrecht:2002uz}, we believe that 
such a dynamical approach offers a much more fundamental understanding
of initial conditions of the universe.  

Interestingly, DKS get results that are very negative for inflation,
and also for big bang 
cosmology in general. According to DKS, inflation is exponentially
less likely than the big bang simply fluctuating into existence without an
inflationary period.  Furthermore, the familiar big bang history for
the observed universe is exponentially less likely than some much more
random fluctuation forming the universe as we see it today. 

Our main goal is to investigate the general issue of
the start of inflation, and particularly the challenges for inflation
raised by the DKS paper (which we argue might reflect a very general
problem \cite{aakitp}).  A key part of this paper is an alternative
calculation of 
the probability that inflation formed our universe.  Our calculation
employs much of the DKS framework, and also takes the meta-universe to
be a fluctuating equilibrium state. Our method is different in the
specifics of how the probabilities are calculated, and represents what
we argue is a more traditional approach (based on reasonably
rigorous semiclassical methods).   Our calculation 
shows (in a quantified and concrete form) that inflation is
exponentially {\em favored} over other histories of our observed
universe.  We also suggest a modest extension of the DKS formalism
that also predicts that the standard Big Bang history of the universe
is favored over the more random versions considered by DKS\footnote{As
  this work was completed, we learned that 
  Guth and Susskind have been considering similar issues. We thank
  A. Guth and L. Susskind for private communications and for copies of
Guth's slides for the ``3rd Northeast String Cosmology Workshop'' (May
14 2004) where some of their ideas were presented}.  

This paper is related to questions about the relationships between
inflation, entropy and the arrow of time which have appeared in one
form or another since the early days of inflation.  Our discussion
allows these issues to take a more quantitative form.  For
completeness we comment further on these connections in Appendix B.

Section \ref{sec:dks} reviews the DKS calculations and results.  We
identify the few key ingredients that lead to problems for inflation
and argue that if one accepts these ingredients the problems for inflation
are likely to persist in a wide variety of different scenarios. 
In section \ref{sec:probsbb} we
discuss the problems faced by the standard big bang in the DKS
picture. We show how a modest extension of the DKS calculations
(introduced in section \ref{sec:xdks}) 
alleviates that particular problem, although we also argue in section
\ref{sec:bb} that the 
problem is replaced with another one that was first explored by
Boltzmann a century ago.
As discussed in \cite{Albrecht:2002uz}, inflation is the first idea with
a chance to resolve the so-called ``Boltzmann's brain paradox'', but
in the extended DKS calculations (which disfavor inflation) the
paradox remains.  

Section 
\ref{sec:tun} presents our own calculation.  We embrace many of the
same assumptions and formalism of DKS, but at a crucial step where DKS
use holographic considerations we use standard semiclassical tunneling
rates from the existing literature. Section \ref{sec:disc} gives further
interpretation and discussion of the two methods. We argue that at the
very least we have constructed a concrete formalism that
reflects the standard intuition about inflation (and also resolves
the Boltzmann's brain paradox). However, we also acknowledge the
strong theoretical basis for the DKS approach based on holography. We
conclude that further investigation contrasting the two methods 
might yield very interesting insights into the nature of quantum
gravity and the early universe, insights which stand to either
validate or destroy key components of modern theoretical
cosmology.

\section{Review of the DKS results}
\label{sec:dks}

\subsection{The general scheme}

Dyson, Kleban and Susskind \cite{Dyson:2002pf} consider the case where the current cosmic
acceleration is given by a fundamental cosmological constant
$\Lambda$.  In that 
picture the universe in the future approaches a de Sitter space, with
a finite region enclosed in a horizon filled with low temperature
Hawking radiation.  
The horizon radius $R_\Lambda$ is given by
\begin{equation}
R_\Lambda \equiv \sqrt{3 \over \Lambda}
\end{equation}
and the Hawking temperature is given by 
\begin{equation}
T_H = {1 \over 2 \pi R_\Lambda} 
\end{equation}
We use  conventions where $\hbar = c = k_B = G \equiv m_P^{-2}
\equiv l_P^2= 1$. With our conventions for $\Lambda$ the equivalent
mass density corresponding to $\Lambda$ is given by $\rho_\Lambda =
\Lambda / 8 \pi G $

DKS take the so-called ``causal patch'' view that, since physics
outside the de Sitter horizon is truly irrelevant to physics inside,
one should consider the physics inside the horizon as the complete
physics of the universe \cite{tom,hks,witten,willylenny,bousso}.  The
points of view of different observers 
that might have different 
horizons should be given by re-arranging (probably
in some highly non-local way) the same fundamental degrees of
freedom, without increasing the total number of degrees of freedom
required to describe the whole universe. 

In this picture, the entire universe is a truly finite system which,
when allowed to evolve sufficiently long will achieve an
equilibrium state, namely the de Sitter space.  The entropy of this
equilibrium state is given by \cite{Gibbons:mu}
\begin{equation}
S_\Lambda =  \pi  R_\Lambda^2 / l_P^2.
\end{equation}
One then has the following picture of the meta-universe:  The
meta-universe is just the finite universe within the causal patch.
The meta-universe spends by far  most of its time in the equilibrium
state: de Sitter space full of Hawking radiation.  This equilibrium
state is constantly fluctuating, and on very rare occasions extremely
large fluctuations occur.  In this picture the universe as we see it
should be regarded as one of the very rare fluctuations out of de
Sitter equilibrium.  Our own destiny is to return to de Sitter equilibrium, a
process that is just starting to become noticeable with the detection
of the cosmic acceleration.  (Note, we are talking about statistical
mechanics here, not thermodynamics, so the entropy will go down just
as often as it will go up as the system fluctuates out of and then
back into equilibrium.) 

DKS assume this system has a sufficient level of ergodicity to use the
following estimate of the probabilities of different fluctuations.
Let $N_T$ be the total number of states available to the system:
\begin{equation}
N_T \equiv e^{S_\Lambda}.
\label{NT}
\end{equation}
Any fluctuation $F$  will start in equilibrium and evolve to some
state with minimum entropy $S_F$, at which point the entropy starts
increasing and the system return to the equilibrium state. 

The ergodic assumption (which says that the system spends roughly an equal
amount of time in each microstate) gives the following probability for a given
fluctuation $F$ in terms its minimum entropy $S_F$:
\begin{equation}
P_F \equiv {N_F \over N_T} \equiv {e^{S_F} \over e^{S_\Lambda}} =
e^{\left( S_F - S_\Lambda \right)}
\label{eqn:pf}
\end{equation}
where $N_F$ is the number of microscopic states with coarse-grained
entropy $S_F$

\subsection{Problems for Inflation}
One can use this picture to compare the probabilities of two different 
types of fluctuations that both lead to the universe we observe
today.  One of these (labeled by $I$) passes through a period of
inflation, the other is simply the ordinary big bang cosmology, with
no inflation (labeled by $BB$). 

For  $S_{BB}$, the minimum entropy of the plain big bang fluctuation,
one can use the entropy of the standard big bang cosmology in the
early universe:
\begin{equation}
S_{BB} \approx 10^{85}.
\label{eqn:sbb}
\end{equation}
Black holes (such as those at the centers of galaxies) dominate the
entropy of the universe today and give larger total value for the
entropy ($S \approx 10^{90}$), but the quantity required for the DKS
calculation is the lower value.   

During cosmic inflation, the universe is dominated by an effective
cosmological constant and looks (for a finite time) very much like de
Sitter space.  DKS estimate the entropy of the universe at that time by the
Gibbons-Hawking entropy of the equivalent de Sitter space:
\begin{equation}
S_I \approx { \left(R_I \over l_P \right)^2} \approx \left( m_P \over
m_I \right)^4 \approx 10^{10}
\label{eqn:si}
\end{equation}
where $ R_I$ is the effective de Sitter radius during inflation and
$m_I$ is the characteristic energy scale of inflation.  (Throughout
this paper when we assign a value to $m_I$ we use $m_I = 10^{-2.5}
m_P$.)

Using Eqn. \ref{eqn:pf} to construct the probabilities (with
Eqns. \ref{eqn:sbb} and 
\ref{eqn:si} for the entropies) gives the following comparison:
\begin{equation}
{P_I \over P_{BB} } = e^{(S_I - S_{BB})} \approx e^{-S_{BB}} \ll 1.
\label{eqn:pibb}
\end{equation}
In this scheme fluctuations that produced the universe we see via
inflation are 
strongly disfavored compared to fluctuations that produce big bang
scenarios without inflation.   

\subsection{The reason for the problem}
\label{sec:reason}
Let us zero in on the origin of this result, which seems exactly the
opposite of the standard intuition about inflation.  

The standard thinking is that the fluctuation required to start
inflation (which after all requires a fluctuation over just a
few Hubble volumes at the inflation scale) is surely much more likely
than a fluctuation that gives rise to the entire big bang universe
directly.  From this perspective, the small entropy of the inflating
state seems to be the key {\em advantage} of the inflationary picture,
while according to DKS, it is the feature that causes inflation to be
strongly disfavored.  

To illustrate the origin of this these dramatically different
perspectives, consider an ordinary box of radiation in equilibrium at
temperature $T$. Consider two possible rare fluctuations.  In the first,
all the radiation in a volume of $ 1 cm^3$ in one corner fluctuates
further into the corner so it only occupies a volume of $ 1 mm^3$. The
second fluctuation is similar, but the initial regions is $ 2 cm^3$
while the final region is still $ 1 mm^3$.  Intuitively, the second
fluctuation is much more rare, even though the entropy of the $ 1
mm^3$ region is larger for the 2nd case.  The reason is that for
fluctuation 1, more of whole system remains in equilibrium during the
fluctuation, making the corresponding state more likely.
Specifically, the entropy in Eqn. \ref{eqn:pf} is the entropy for the {\em
  entire} system which is larger for fluctuation $1$ because more of
the system remains in equilibrium.  Using entropy density $ \approx
5 \times 10^8 cm^{-3}$ for a photon gas at room temperature and
$S_t$ for the total equilibrium entropy one can
evaluate Eqn. \ref{eqn:pf} to get 
\begin{equation}
{P_1 \over P_2 } = e^{(S_1 - S_2)} \approx e^{(S_t - 5 
\times 10^8) - (S_t -
    10^9)}  =  e^{ 5\times 10^8} \gg1
\end{equation}
which quantifies the intuitive result that fluctuation $1$ is more
likely.  (The positive contribution from the entropy of the gas in the
$1mm^3$ region is completely subdominant.)

So if inflation requires only a few inflation era Hubble volumes to
get started, why is not $S_I \approx S_\Lambda$?  Surely while one
little region starts inflating, the rest of the universe is free, at
least at first, to be doing whatever it likes (which would mean
staying in equilibrium). Why does not that mean that inflation is
strongly favored over other paths to the big bang that have a
larger part of the entire system participate in the fluctuation to begin with?

DKS use a very small value for $S_I$ because of the principles of
causal patch physics which they employ.  Because a horizon forms
during the inflationary period, these principles dictate that an
observer inside the horizon sees {\em all} the degrees of freedom of
the universe inside the horizon with him.  From his point of view
there simply is no ``outside the horizon'', and $S_I$ must be
evaluated using only what this observer sees.  It is exactly this feature
of their analysis that turns what might seem like the main advantage
of inflation (the simplicity of the initial fluctuation) into a
extremely serious liability.

\subsection{An extension of DKS}
\label{sec:xdks}

The formation of the horizon is crucial to DKS's evaluation of
$S_I$. However the path to the SBB that does {\em not } include
inflation is usually not thought of as forming a horizon and since $
S_{BB} \ll S_\Lambda $ one might think of this fluctuation, at its minimum
entropy state, as a small localized perturbation on the de Sitter
meta-universe.  In the far-field limit any such localized perturbation
will have a Schwarzschild geometry, and 
Gibbons and Hawking~\cite{Gibbons:mu}
showed in this situation the Schwarzschild perturbation changes the
area of the de Sitter horizon according to
\begin{equation}
R_\Lambda^2 \rightarrow R_\Lambda^2 - R_\Lambda\, R_S\,,
\label{eqn:ghrlrs}
\end{equation}
where $R_S$ is the Schwarzschild radius of the perturbation.  This
suggests that an improved estimate of $S_{BB}$ might
be~\footnote{We are assuming here that the Schwarzschild
  radius outside 
  this perturbation is given by $l_P \sqrt{S_{BB}}$. This may not be
  exactly right, but we expect that corrections to this are
  unlikely to change the 
  qualitative result. }
\begin{equation}
\tilde{S}_{BB} =  S_\Lambda - \sqrt{S_\Lambda S_{BB}} + S_{BB} \approx
S_\Lambda - \sqrt{S_\Lambda S_{BB}} 
\label{eqn:sbbtilde}
\end{equation}
Using $\tilde{S}_{BB}$ in Eqn. \ref{eqn:pibb} gives
\begin{equation}
{P_I \over P_{BB} } = e^{(S_I - \tilde{S}_{BB})} \approx e^{- S_\Lambda
  + \sqrt{S_\Lambda S_{BB}} } \approx e^{- S_\Lambda
  } \ll 1. 
\label{eqn:pibbtilde}
\end{equation}
which even more strongly disfavors inflation.  Here, in the absence of
a horizon within the $BB$ fluctuation, we have allowed the counting of
entropy for the $BB$  fluctuation to include the ``outside'' part of
the meta-universe. The $BB$ fluctuation has gained further ground compared
with Eqn. \ref{eqn:pibb} by the recognition that the $BB$ fluctuation
is small and allows most of the meta-universe to remain in an
equilibrium state.  That increases the total entropy associated with
the fluctuation, and thus increases its probability.

This extension of the DKS calculation also lets one express the
regular intuition about inflation in the following way:  If one
forgets about the principles of causal patch physics and just forges
ahead treating the inflationary fluctuation in a similar manner to
the $BB $ fluctuation, one might construct  
\begin{equation}
\tilde{S}_{I} =  S_\Lambda - \sqrt{S_\Lambda S_{I}} + S_{I} \approx
S_\Lambda - \sqrt{S_\Lambda S_{I}} 
\label{eqn:sitilde}
\end{equation}
which would lead to 
\begin{eqnarray}
{P_I \over P_{BB} } = e^{(\tilde{S}_I - \tilde{S}_{BB})}&& \approx
  \left(e^{-\sqrt{S_I} + \sqrt{ S_{BB}} } \right)^{\sqrt{ S_\Lambda}}
\approx\nonumber\\
&&\approx  \left(e^{\sqrt{ S_{BB}} } \right)^{\sqrt{ S_\Lambda}}
  \gg 1. 
\label{eqn:pibb2tilde}
\end{eqnarray}
In this expression inflation gets credit for the small entropy of the
inflating region, in that the small value of $S_I$ allows more of the
rest of the universe to remain in equilibrium. This allows the total
entropy of the system during an inflationary fluctuation to be
larger, assigning it a greater probability. Equation
\ref{eqn:pibb2tilde} expresses the standard intuition about inflation
but violates the principles of causal patch physics.  We will develop
a more carefully constructed expression for $P_I$ that has similar
features in Section \ref{sec:tun}.

\subsection{The generality of the problem for inflation}
\label{sec:generality}

It is tempting to try and view the failure of inflation in the DKS
picture as a result of other assumptions and details of their
calculation.  In particular, in the DKS picture the finiteness of the 
whole
meta-universe imposed by the late time de Sitter horizon $R_\Lambda$
appears to exclude the possibility of eternal inflation (at least as
it is traditionally understood).  In eternal inflation \cite{eternal},
inflation starts with some fluctuation and then continues eternally
into the future, seeding additional inflating regions via quantum
fluctuations.  Also, infinitely large numbers of regions 
stop inflating and
reheat to produced ``SBB'' regions that look like the universe we
observe. It seems reasonable to argue that in this picture the infinite
numbers of SBB regions will overwhelm any suppression of the
probability to start inflation and allow inflation to win any
competition with other channels for producing SBB regions.  

However,
as long as the principles of causal patch physics require one to
assign very small values to $S_I$, it is far from clear that eternal
inflation can resolve the problem.  As one allows the size of the
meta-universe to diverge in order to accommodate eternal inflation it is
quite possible that the $P_I/P_{BB}$ will go to zero fast enough that
inflation never wins, despite the increasingly large numbers of SBB
regions produced by inflation.  

For example one can adapt Eqn. \ref{eqn:pibbtilde} to this situation
by thinking of $\Lambda$ not as the source of cosmic acceleration
today (which can be provided by quintessence) but simply as a
regulator that allows one to define the meta-universe in a concrete
way.  If one lets $ \Lambda \rightarrow 0$ the size of the
meta-universe $R_\Lambda$ will diverge, allowing more room for eternal
inflation, but $S_\Lambda$ will also diverge, driving $P_I/P_{BB}
\rightarrow 0$.  In this analysis taking  $ \Lambda \rightarrow 0$
only increases the problem for inflation, since to start inflation one
now has to cause a divergently large universe to fluctuate into a
region with finite entropy.  (The divergent ``volume factors'' from
eternal inflation that enhance the probability of producing the
observed universe via inflation only appear as an 
inverse power of $\Lambda$ in
the prefactor and are unable to compensate for the huge exponential
suppression.) 

Of course, there are probably other ways of taking the infinite
universe limit. Our point here is that the infinite universe limit
(whether in the context of eternal inflation or more general
considerations such as the ``string theory
landscape''~\cite{Susskind:2003kw}) is 
not a sure way to save inflation. The  causal patch arguments that
assign low entropy to the {\em whole universe} when there exists just
a single inflating patch can create even bigger
problems for larger meta-universes.  At the very best, this limit throws
inflation at the mercy of problematic debates about defining measures
and probabilities  for  infinite systems.

\section{The problem for the Standard Big Bang}
\label{sec:probsbb}

\subsection{The problem according to DKS}
\label{sec:probsbbdks}

The DKS calculations do not just create problems for inflation.  DKS
consider variations to the SBB which increase $S_{BB}$ to some new
value we will designate by $S_{B2} > S_{BB}$.  This could be a version of the
big bang, for example, with a somewhat higher value for the temperature
of the cosmic microwave background today.  Applying the DKS
scheme one gets
\begin{equation}
{P_{B2} \over P_{BB} } = e^{({S}_{B2} - {S}_{BB})} > 1 
\label{eqn:pb2bb}
\end{equation}
which favors the modified big bang scenario.  Certainly the $B2$
fluctuation requires some strange out-of-equilibrium behavior in the
early universe, in contrast to the $BB$ fluctuation.  That does not
mean much however, because in this
scheme the big picture is that {\em anything} that looks at all like
the SBB is an out-of-equilibrium fluctuation of the
meta-universe. Our job as cosmologists is to make predictions based on
the most likely fluctuation to create what we see.  Equation
\ref{eqn:pb2bb} is interpreted by DKS as a (failed) prediction that our
universe should be found in a higher entropy state than we actually
observe.

\subsection{Extended DKS solves the SBB problem}
\label{sec:probsbbxdks}

One can also apply the extended DKS formalism of section
\ref{sec:xdks} to the comparison of the $B2$ and $BB$ fluctuations
discussed above, giving
\begin{equation}
{P_{B2} \over P_{BB} } = e^{(\tilde{S}_{B2} - \tilde{S}_{BB})} \approx
  \left(e^{-\sqrt{S_{B2}} + \sqrt{ S_{BB}} } \right)^{\sqrt{ S_\Lambda}}
  \ll 1. 
\label{eqn:pb2bbtilde}
\end{equation}
The extended DKS scheme reverses the fortunes of the standard Big Bang
{vs.} other $B2$ type fluctuations with higher entropy. The
reason for this reversal is that in the extended DKS scheme larger
values of $S_{B2}$ mean more of the meta-universe is tied up in
creating the variant fluctuation $B2$ and is thus removed from 
equilibrium.  The corresponding entropy reduction in the meta-universe
($\sqrt{S_\Lambda S_{B2}}$) is much greater than the entropy added
back in by the larger value of $S_{B2}$, so the total entropy of the system
for the $B2$  fluctuation ($\tilde{S}_{B2}$) is lower than for the
$BB$ case. Of course in this picture the big bang gets serious
competition from scenarios with $S_{B2} < S_{BB}$.  That topic is
addressed (in an extreme limit) in the next subsection.

\subsection{Boltzmann's Brain}
\label{sec:bb}

A century ago Boltzmann considered a ``cosmology'' where the observed
universe should be regarded as a rare fluctuation out of some
equilibrium state.  The prediction of this point of view, quite
generically, is that we live in a universe which maximizes the total
entropy of the system consistent with existing observations. Other
universes simply occur as much more rare fluctuations.   This means as
much as possible of the system should be found in
equilibrium as often as possible. 

From this point of view, it is very surprising that we
find the universe around us in such a low entropy state. In fact, the
logical conclusion of this line of reasoning is utterly
solipsistic. The most likely fluctuation consistent with everything
you know is simply your brain (complete with ``memories'' of the
Hubble Deep fields, WMAP data, etc) fluctuating briefly out of chaos
and then immediately equilibrating back into chaos again.  This is
sometimes called the ``Boltzmann's Brain'' paradox \cite{bt86}.   The DKS
formalism (extended or otherwise) certainly manifests the Boltzmann's
Brain paradox because it attaches higher probabilities to larger
entropy fluctuation.

As discussed in \cite{Albrecht:2002uz}, cosmic inflation is the only idea we are
aware of that could potentially resolve this paradox. In models where
inflation is the preferred route to the observed universe many brains appear
in a single inflated region, so the probability per brain could be
significantly reduced. Also the brains produced via inflation come
correlated with bodies, fellow creatures, planets, large flat
universes with CMB photons etc.  A much more realistic picture. But
the DKS formalism cannot exploit inflation to resolve the Boltzmann's
Brain paradox because inflation itself is so strongly disfavored in
that formalism.

\section{The semiclassical calculation}
\label{sec:tun}

In this section we construct an alternative calculation of the
probability for a region to start inflating in the de Sitter
meta-universe.

There is a large body of literature addressing the formation of an
inflating region from a non-inflating
state~\cite{Farhi:1986ty,FMP,Farhi:1989yr}.   
It has been well
established that no classical solution can evolve into an inflating
region, but that it {\em is} possible for certain classical solutions
to quantum tunnel into an inflating solution.  Here we apply these
results to process of forming an inflating region in the de Sitter
meta-universe described by DKS. 

We apply the formalism of Fischler, Morgan and Polchinski
(FMP) \cite{FMP} and use their notation.  Farhi et al
\cite{Farhi:1989yr} achieve equivalent results using functional
methods, but we focus on the FMP work because their Hamiltonian
formalism is free of the ambiguities of the functional methods noted in
\cite{Farhi:1989yr}.  FMP consider 
solutions with spherical symmetry and assume an inflaton exists with a
suitable potential to produce inflation.  They also assume that
solutions with regions up and down the potential can be treated in the
thin wall approximation.  The quantum tunneling probability from the inflating
to the non-inflating state is given by
\begin{equation}
P_q \approx e^{-2F}.
\label{eqn:pq}
\end{equation}
FMP do not calculate the prefactors to the exponential, and we do not
require them here for the very broad issues at hand.  The form of $F$
is discussed in detail in Appendix A, where we show that for our
purposes $F$ can be extremely well approximated by 
\begin{equation}
F = { \pi \over 2 l_P^2}  R_I^2  \approx {1 \over
  2} S_I.
\label{eqn:f}
\end{equation}
Here $R_I$ is the de Sitter radius of the inflating region and $R_S$
is the Schwarzschild radius corresponding to the classical solution
that tunnels into the inflating state. 

But $P_q$ gives the probability of tunneling into inflation from a
very specific classical state.  The total probability for starting
inflation by this path will take the form
\begin{equation}
P_I = P_c P_q
\label{eqn:pcpq}
\end{equation}
where $P_c$ is the probability of forming the classical state used to
calculate $P_q$. 

To determine $P_c$ we use the methods of DKS and write
\begin{equation}
P_c = e^{(S_c - S_\Lambda)}
\label{eqn:pcsc}
\end{equation}
where $S_c$ is the entropy of the de Sitter universe in the presence
of the classical solution in question.  As discussed in section
\ref{sec:xdks} Gibbons and Hawking
have shown that $S_c$ is dominated by shrinkage of the de Sitter horizon
\begin{equation}
S_c \approx S_\Lambda - \sqrt{S_\Lambda S_S}
\label{eqn:sclc}
\end{equation}
where $S_S = \pi m_P^2 R_S^2$ taking $R_S$ to be the Schwarzschild
radius of the perturbation to the de Sitter space.
Combining the above results gives
\begin{equation}
P_c =  e^{ - \sqrt{S_\Lambda S_S}}.
\label{eqn:pclc}
\end{equation}
and 
\begin{equation}
P_I = e^{
\left(
 - \sqrt{S_\Lambda S_S}
 + S_I
 - S_S 
\right) } \approx
e^{\left( - \sqrt{S_\Lambda S_S} + S_I  \right) }.
\label{eqn:pi}
\end{equation}
This expression depends on the mass of the classical solution that
tunnels through to inflation via the entropy $S_S$, and is maximized in
the limit $S_S \rightarrow 0$ (vanishing mass).  

The mass $ \rightarrow 0$ limit is an intriguing one, in that is seems
to represent the formation of an inflating region ``from nothing''.
We proceed with caution here, however, since we expect various aspects
of our calculation (such as the thin wall limit and semiclassical
gravity) to break down in zero mass limit.  We take our formula to be
valid down to some lower cutoff value of $S_S$ given by $S_l $.  If $S_l$ is
set by the breakdown of the thin wall approximation, perhaps $S_l
\approx (m_P / m_I )^2 \approx 10^5$.  Perhaps our formula works all
the way down to the Planck scale and $S_l \approx 1$.  The actual
value of $S_l$ is completely irrelevant for our main points (even $S_l
= 0$ is fine).  

We now compare $P_I$ and $P_{BB}$ using extended DKS for $P_{BB}$ and
Eqn. \ref{eqn:pi} for $P_I$:
\begin{equation}
{P_I \over P_{BB} }  \approx
  \left(e^{-\sqrt{S_l} + \sqrt{ S_{BB}} } \right)^{\sqrt{ S_\Lambda}}
\approx
  \left(e^{\sqrt{ S_{BB}} } \right)^{\sqrt{ S_\Lambda}}
  \gg 1. 
\label{eqn:pibbtun}
\end{equation}
Instead of following the causal patch principles this calculation uses
conventional semiclassical methods.  This difference allows the
(barely perturbed) entropy of the de Sitter equilibrium to be included
in calculation of the production rate of inflationary fluctuations.
Our scheme realizes the standard intuition about inflation and
strongly favors inflation over other paths to the universe we
observe. 

\section{Discussion and Conclusions} 
\label{sec:disc}

We have argued that a meta-universe picture, in which inflation 
competes in a direct and quantifiable way  with other cosmological
scenarios, is crucial to validating the expectations that inflation is
a ``more likely'' or ``more natural'' origin of our observed
universe. 

The methods of Dyson Kleban and Susskind gave the most concrete
picture yet of a meta-universe which allows one to quantify the
competition between different cosmologies, but the results of this
competition are completely reversed from the expectations of most
cosmologists. According to DKS inflation is exponentially less
probable than big bang scenarios without inflation, and 
variants of the big bang which have a higher entropy for the observed
universe are exponentially favored over the big bang 
scenario itself.   

In this paper we have introduced alternative calculations which, while
very much in the DKS spirit, are sufficiently different that the order
of preference is reversed:  In our calculations inflation is
exponentially favored over an inflation-free big bang, which itself is
favored over the variants of the big bang that beat inflation in the
DKS calculation.

The most important difference between our methods and those of DKS is
the role played by the principles of causal patch physics.  The causal
patch rules state that once a horizon forms in an inflating region
the ``entire universe'' is inside the horizon.  The region ``outside''
the inflating region is not represented by different degrees of
freedom, but is supposed to be described by the same degrees of
freedom re-expressed in terms of different variables to account for
the different observers.  This feature is at the heart of the negative
results for inflation from DKS.  Specifically, it is the use of the
entropy inside the horizon of the inflating region along with ergodic
arguments that harms inflation in their scheme. We argue that any theory that 
follows these rules is likely to disfavor inflation even
if other aspects of the theory differ greatly from the DKS scheme (by
including, for example eternal inflation or a large string theory landscape).

Our calculation does not follow the specific causal patch rules of
DKS.  Instead we 
view the formation of an inflating region as a quantum
tunneling event.  We calculate tunneling rates based on well
established semiclassical (``WKB'') methods for tunneling through a classically
forbidden region, which one can hope would not 
get significant corrections from a deeper theory of quantum
gravity. From the point of view of this paper, the key aspect the
semiclassical quantum tunneling problem is that the different
sides of the classically forbidden region are described by 
{\em different} states in the same Hilbert space. The tunneling process
describes the flow of quantum probability from one side of the barrier
to the other, and describes a global state of the entire system.  This
perspective seems to be in marked contrast to the causal patch view that says
the what we view semiclassically as ``two sides of the barrier'' are
not actually represented by different 
parts of the space of states. Instead, the ``fluctuating toward
inflation'' state and the inflating state are seen as
re-parameterizations of the same state in the same space.  This
difference is at the heart of the sharply differing results from the
two methods

We feel that the reconciliation of the these two methods presents a very
interesting problem in quantum gravity and cosmology.  Perhaps deeper
insights into quantum gravity will show us that at least one of the
approaches is simply wrong.  Another interesting possibility is that
the one or both of these schemes require a more careful
implementation (for example a refinement of the ergodic arguments)
that will actually bring the two approaches into 
quantitative agreement.  Whatever the outcome, it appears that the
viability of cosmic inflation theory hangs in the balance.  Different
outcomes could either enhance or end inflation's prominence as a theory
of the origins of the universe.

\begin{acknowledgments}
We thank T. Banks, N. Kaloper, M. Kaplinghat, M. Kleban, N. Turok and
L. Susskind for helpful discussions. One of us (A.A.) also thanks the
{\em Nobel Symposium on String Theory and Cosmology} and the {\em KITP
  Superstring Cosmology Workshop} where many of these discussions took
place. This work was supported in part by DOE grant DE-FG03-91ER40674. 
\end{acknowledgments} 

\appendix

\section{Calculating Tunneling Rates}

Fischler Morgan and Polchiski consider
non-inflating classical solutions that quantum tunnel to inflating
classical solutions. The solutions are spherically symmetric and they
use a thin wall approximation where the stress-energy is zero outside
of some region, and has a cosmological constant $\equiv \Lambda_I$
inside.  The regions are separated by a spherical wall with tension
$\mu$.  In the outside region the spacetime is Schwarzschild with mass
$M$.  FMP use semiclassical Hamiltonian methods which are described in
detail in \cite{FMP} and references therein\footnote{Technically FMP
  discuss tunneling out of solutions in Minkowski space whereas we
  consider tunneling from a background cosmology with a cosmological
  constant $\Lambda$.  However, since $\Lambda \ll \Lambda_I$  we
  expect correction due to $\Lambda \neq 0$ to be inconsequential for
  the tunneling calculation.
}.  Although the actual classical solutions that tunnel into inflation
start with a singularity, FMP discuss how in a more complete treatment
these solutions could emerge from excitations other than a
singularity.  In our case we think of these solutions fluctuating out of
the thermal Hawking radiation of de Sitter space.
  Their tunneling
probability is given by  
\begin{equation}
P \propto e^{-2F}
\label{eqn:gamma}
\end{equation}
where
\begin{equation}
F = F_I + F_O + \hat{F}
\label{eqn:fsum}
\end{equation}
and
\begin{eqnarray}
&&F_I
+F_O
=\left\{
\begin{array}{l} \frac{\pi}{2\,G}\,\left(R_2^2-
R_1^2\right),\quad M_{cr}>M>M_S\\ \frac{\pi}{2\,G}\,\left(R_2^2-
R_S^2\right),\quad M_D<M<M_S\\ \frac{\pi}{2\,G}\,\left(R_I^2-
R_S^2\right),\quad M<M_D \end{array} \right.\nonumber\\ &&.
\label{eqn:fio}
\end{eqnarray}
Our $R_I$ is FMP's $R_d$ (the de Sitter radius during inflation), and
$R_S = 2GM$. The values of the transverse radius at the classical
turning points between which the tunneling occurs are given by $R_1$
and $R_2$. The third term in $F$ is
\begin{widetext}
\begin{equation}
\hat F = { R_I^2 \over G \lambda} \int_{\phi _1 }^{\phi _2 } {\hat \phi }\,\left[ {\arccos \left( {\frac{{\left( {\psi _S /\hat \phi ^2
} \right) - \hat \phi \left( {1 - \frac{1}
{\lambda }} \right)}}
{{2\sqrt {1 - \left( {\hat \phi ^2 /\lambda } \right)} }}}
\right) - \arccos \left( {\frac{{\left( {\psi _S /\hat \phi
^2 } \right) - \hat \phi \left( {1 + \frac{1}
{\lambda }} \right)}}
{{2\sqrt {1 - \left( {\psi _S /\hat \phi \lambda } \right)}
}}} \right)} \right]\,d\hat\phi.
\label{eqn:fhat}
\end{equation}
\end{widetext}
Here we use the rescaled variables 
\begin{equation}
\phi \equiv {R \over R_I}\sqrt{\lambda}; \quad \psi_S \equiv {R_S \over R_I
}\lambda^{3/2}.
\label{eqn:phipsidef}
\end{equation}
The turning points $\phi_1$ and $\phi_2$ are the roots of 
\begin{equation}
\lambda \left[ {\left( {\frac{{\psi _S }}
{{\hat \phi ^3 }} - 1} \right) - \frac{1}
{\lambda }} \right]^2  + 4\left( {\frac{{\psi _S }}
{{\hat \phi ^3 }} - \frac{\lambda }
{{\hat \phi ^2 }}} \right) = 0
\label{eqn:pizero}
\end{equation}
where
\begin{equation}
\lambda = { \Lambda_I \over 3 G^2\mu^2} \approx \left(\frac{m_P}{m_I}\right)^2.
\label{eqn:lambda}
\end{equation}
The mass scales $M_D$ and $M_S$ in Eqn.~\ref{eqn:fio} are worked out in
\cite{bgg} to be:
\begin{equation}
M_S = \bar{M}\left(1 - {1 \over 4}\gamma^2 \right)^{1/2} =
\bar{M}\left( 1 + O(1/\lambda)\right)
\label{eqn:ms}
\end{equation}
and
\begin{equation}
M_D = \bar{M}{ 1 - {1 \over 2}\gamma^2  \over 1 - {1 \over 4}\gamma^2 }=
\bar{M}\left( 1 + O(1/\lambda)\right)
\label{eqn:md}
\end{equation}
where
\begin{equation}
\gamma \equiv { 2 \over \sqrt{ 1 + \lambda} }
\label{eqn:gammadef}
\end{equation}
and $\bar{M} \equiv m_P^2R_I/2$.

We use the quantity $F$ in Eqn. \ref{eqn:pcpq} to give $P_I = P_ce^{-2F}$.
The classical part $P_c$ is of a form that maximises $P_I$ in the
$\psi_s \rightarrow 0$ (small $M$) limit, so we want to evaluate $F$ in
this limit. This gives
\begin{equation}
F_I +F_O \rightarrow {\pi \over 2G} R_I^2 \approx \left( { m_P \over
  m_I }\right)^4
\label{eqn:fiom0}
\end{equation}
and 
\begin{eqnarray}
\hat{F} \rightarrow &&- \frac{\pi}{2} \left( \frac{R_I}{l_P}\right)^2
\frac{1}{\lambda^2}
\times ( 1 + O[\lambda^{-1}] ) \approx \nonumber\\
&& \approx - \frac{\pi}{2} \left( { 3 \over 8 \pi}
\right)^3 \ll F_I + F_O
\label{eqn:fhtlim}
\end{eqnarray}
(where the last approximation assumes $\mu \approx m_I^3$).
So in this limit the first part of $F$ dominates and we can take 
\begin{equation}
F = {\pi \over 2G} R_I^2
\label{eqn:fm0}
\end{equation}
to an excellent approximation.

\begin{figure}
\includegraphics[width=2.5in]{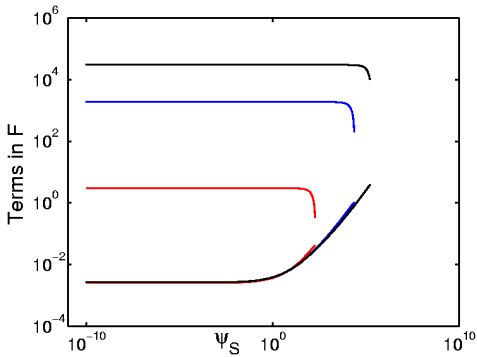}
	\caption{
		\label{fig:fps}
The values of $F_I + F_O$ (top three curves) and $\hat{F}$ (the highly
overlapping bottom curves)
as a function of the rescaled mass parameter $\psi_S$ over the range
$10^{-10} < \psi_S < \psi_{cr}$.The three values of the pair
($m_I/m_P$,$\lambda$) shown here are given by ($0.5$,$33.5$),
($0.1$,$838$), and ($0.05$,$3350$) in order of increasing
$\psi_{cr}$.   The main point of this figure is that $F_I + F_O \gg
\hat{F}$ for all relevant values of $\psi_S$ for any realistic value
of $\lambda$.  More detailed features of this plot are discussed in the text.
}
\end{figure}

In \cite{bgg} it is also shown that the turning point solutions do not
exist for $M > M_{cr}$.  In that regime there is no tunneling and no
chance to produce inflation.  Like $M_S$ and $M_D$, $M_{cr} $ also
takes the form $M_{cr} = \bar{M}(1 + O(1/\lambda))$.  For realistic
models of inflation $m_I \leq O(10^{-2.5})m_P$, so again taking $\mu
\approx m_I^3$, $1/\lambda \ll 1$.  Thus for realistic models, $M <
M_D$ holds for all $M$ except for a tiny range $\Delta M \approx
\bar{M}/\lambda$ near the maximum value $M_{cr} \approx \bar{M}$.
The form of $P_c$ dictates that {\em minimum } values of $M$ are
the relevant ones, so we can use the $M < M_D$ part of
Eqn. \ref{eqn:fio} for $F_I + F_O$ in $P_q$ for all values of $M$
without producing any significant errors.

Figure \ref{fig:fps} shows $F_I + F_O$ and $\hat{F}$ for three
different values of 
$m_I$ (and $\lambda$, which we take to be specified uniquely from $m_I$
by using $\mu = m_I^3$ in Eqn. \ref{eqn:lambda}).  We've
chosen unrealistically large values of $m_I$ so that key 
features can be shown more easily on the plot.  The pair of curves
corresponding to each value of $m_I$ extends all the way to the
maximal value
of $\psi_S = \psi_{cr} \equiv \psi_S(M_{cr}) = \lambda^{3/2}(1 + O(1/\lambda))$
corresponding to the given value of $m_I$ .  

We see that the $M\rightarrow 0$ limit is a good approximation for
$\hat{F}$ for values of $\psi_S$ up to $\psi_S \approx 1$.  Above
$\psi_S = 1$, $\hat{F}$ increases, but remains orders of magnitude
smaller than  $F_I + F_O$ except possibly in the tiny (unresolved)
region near the maximal value $\psi_{cr}$.  Note that the $\hat{F}$ curves
coincide (over their defined ranges) except for corrections
$O(1/\lambda)$ which are barely visible on this plot due to the
(large) chosen values of $\lambda$.

As discussed in Section \ref{sec:tun}, the most significant values of
$P_q$ are those corresponding to masses given by the cutoff value
$m_l$ (which we expressed in terms of the corresponding black hole
entropy $S_l$).  The corresponding cutoff value of $\psi_S$ is given
by $ \psi_l = \sqrt{S_l /S_I}\lambda^{3/2} \approx \sqrt{S_S
/S_I}\psi_{cr}$.  One can see from Fig. \ref{fig:fps} that with the
possible exception of values of $\psi_l$ extremely close to
$\psi_{cr}$ (over a region too narrow to resolve on this plot),  $F
\approx F_I + F_O$ will be an excellent approximation 
for the purposes of our calculations. Since the maximum
value of the cutoff proposed here gives $S_l = \sqrt{S_I}$, $ \psi_l
\ll \psi_{cr}$ so we are always considering values well away from the
narrow $\psi_S \approx \psi_{cr}$ zone. Even at the closest approach
shown $\hat{F} \ll F_I + F_O$, and the gap widens with decreasing
$m_I$.  Thus throughout this paper we take
\begin{equation}
F = F_I + F_O = \frac{\pi}{2\,G}\,\left(R_I^2- 
R_S^2\right) \approx \frac{\pi}{2\,G}R_I^2.
\label{eqn:fioapprox}
\end{equation}
~in $P_q$.

\section{Relationship to issues raised by Penrose and others}
There is some connection between our discussion here and conceptual
issues that have been discussed over some time in connection with
inflation, especially the work of Page \cite{Page:1983uh} (in
responding to Davies 
\cite{Davies:1983nf}) and later Penrose \cite{Penrose}, Unruh
\cite{Unruh} and Hollands and Wald \cite{Hollands:2002xi}.  Page and Penrose 
emphasize the point that initial conditions which given the big bang a
thermodynamic arrow of time must necessarily be low 
entropy and therefore ``rare''.  There is no way the initial conditions can
be typical, or there would be no arrow of time, and this fact must
apply to inflation and prevent it from representing ``completely
generic'' initial conditions. 

The position we take 
here (which was suggested by Davies in \cite{Davies:1984qc} and is the same one
taken by DKS and emphasized at length in 
\cite{Albrecht:2002uz}) is basic acceptance of this point.  If you
can regard the big bang as a fluctuation in a larger system it must
be an exceedingly rare one to account for the observed thermodynamic
arrow of time.  Also, we believe that this is the most attractive
possibility for a theory of initial conditions. Other theories of
initial conditions seem to us more ad hoc, and less compelling. 

There is an additional point that appears in \cite{Unruh} and
\cite{Hollands:2002xi}, but which many 
(including one of us, AA) recall also being discussed orally (but
apparently not in print) by Penrose: It might be argued that
inflation, which has a lower entropy initial state than the big bang
must necessarily be more rare than a fluctuation giving a big bang
without inflation.  For a number of reasons this point of view never
really caught on.  One reason is that intuitively it seemed likely that
a careful accounting of degrees of freedom outside the observed
universe would reverse that conclusion.  Hollands and Wald
specifically note this view, although they also seem drawn to the Penrose
argument. 

All these issues play out in this paper, but in a more concrete form. DKS
have a specific reason why they ignore the external degrees of freedom
(there aren't any separate external degrees of freedom in the causal
patch analysis). DKS are able to 
quantify the serious problems this causes for inflation.  Our calculation
explicitly {\em does} account for external degrees of freedom and we
show quantitatively that that change does indeed reverse the fortunes
of inflation.

\bibliography{recur.bib}

\end{document}